\documentclass{aa}  

\usepackage{graphicx}
\usepackage{txfonts}
\usepackage{xcolor}
\usepackage{hyperref}
\hypersetup{colorlinks, linkcolor=blue, citecolor=blue, urlcolor=blue}
\usepackage{amsmath}
\usepackage{ulem}
\usepackage{soul}

\newcommand{\te}{t_{\rm E}}
\newcommand{\thetae}{\theta_{\rm E}}

\newcommand{\dl}{D_{\rm L}}

\definecolor{brown}{rgb}{0.59, 0.29, 0.0}
\definecolor{darkgreen}{rgb}{0.0, 0.42, 0.24}
\definecolor{darkblue}{rgb}{0.01, 0.31, 0.59}
\definecolor{darkblue}{rgb}{0.0, 0.25, 0.42}
\definecolor{blue}{rgb}{0.0,0.0,1.0}
\definecolor{green}{rgb}{0.0,1.0,0.0}

\begin{document}

\title{KMT-2019-BLG-0797: binary-lensing event occurring on a binary stellar system}

\author{
     Cheongho~Han\inst{1} 
\and Chung-Uk~Lee\inst{2} 
\and Yoon-Hyun~Ryu\inst{2} 
\and Doeon~Kim\inst{1}
\\
(Leading authors),
\\
     Michael~D.~Albrow\inst{3}   
\and Sun-Ju~Chung\inst{2,4}      
\and Andrew~Gould\inst{5,6}      
\and Kyu-Ha~Hwang\inst{2} 
\and Youn~Kil~Jung\inst{2} 
\and Hyoun-Woo~Kim\inst{2} 
\and In-Gu~Shin\inst{2} 
\and Yossi~Shvartzvald\inst{7}  
\and Jennifer~C.~Yee\inst{8}    
\and Weicheng~Zang\inst{9}     
\and Sang-Mok~Cha\inst{2,10} 
\and Dong-Jin~Kim\inst{2} 
\and Seung-Lee~Kim\inst{2,4} 
\and Dong-Joo~Lee\inst{2} 
\and Yongseok~Lee\inst{2,10} 
\and Byeong-Gon~Park\inst{2,4} 
\and Richard~W.~Pogge\inst{6}
\\
(The KMTNet Collaboration),\\
}

\institute{
     Department of Physics, Chungbuk National University, Cheongju 28644, Republic of Korea  \\ \email{\color{blue} cheongho@astroph.chungbuk.ac.kr}     
\and Korea Astronomy and Space Science Institute, Daejon 34055, Republic of Korea                                                                        
\and University of Canterbury, Department of Physics and Astronomy, Private Bag 4800, Christchurch 8020, New Zealand                                     
\and Korea University of Science and Technology, 217 Gajeong-ro, Yuseong-gu, Daejeon, 34113, Republic of Korea                                           
\and Max Planck Institute for Astronomy, K\"onigstuhl 17, D-69117 Heidelberg, Germany                                                                    
\and Department of Astronomy, The Ohio State University, 140 W. 18th Ave., Columbus, OH 43210, USA                                                       
\and Department of Particle Physics and Astrophysics, Weizmann Institute of Science, Rehovot 76100, Israel                                               
\and Center for Astrophysics $|$ Harvard \& Smithsonian 60 Garden St., Cambridge, MA 02138, USA                                                          
\and Department of Astronomy, Tsinghua University, Beijing 100084, China                                            
\and School of Space Research, Kyung Hee University, Yongin, Kyeonggi 17104, Republic of Korea                                                           
}
\date{Received ; accepted}

\abstract
{}
{
We analyze the microlensing event KMT-2019-BLG-0797. The light curve of the event exhibits two
anomalous features from a single-lens single-source model, and we aim to reveal the nature of the
anomaly.
}
{
It is found that a model with two lenses plus a single source (2L1S model) can explain one feature 
of the anomaly, but the other feature cannot be explained.  We test various models and find that 
both anomalous features can be explained by introducing an extra source to a 2L1S model (2L2S 
model), making the event the third confirmed case of a 2L2S event, following on MOA-2010-BLG-117 
and OGLE-2016-BLG-1003.  It is estimated that the extra source comprises $\sim 4\%$ of the $I$-band 
flux from the primary source.
}
{
Interpreting the event is subject to a close--wide degeneracy.  According to the close solution, 
the lens is a binary consisting of two brown dwarfs with masses $(M_1, M_2)\sim (0.034, 0.021)~M_\odot$, 
and it is located at a distance of $\dl\sim 8.2$~kpc.  According to the wide solution, on the other 
hand, the lens is composed of an object at the star/brown-dwarf boundary and an M dwarf with masses 
$(M_1, M_2)\sim (0.06, 0.33)~M_\odot$ located at $\dl\sim 7.7$~kpc.  The source is composed of a 
late-G-dwarf/early-K-dwarf primary and an early-to-mid M-dwarf companion.  
}
{}


\maketitle

\begin{table*}[htb]
\small
\caption{Microlensing events with more than four bodies \label{table:one}}
\begin{tabular}{lll}
\hline\hline
\multicolumn{1}{c}{Model}                           &
\multicolumn{1}{c}{Event}                           &
\multicolumn{1}{c}{Reference}                       \\
\hline
3L1S                 &  OGLE-2008-BLG-092    &  \citet{Poleski2014}  \\
(planet in binary)   &  OGLE-2007-BLG-349    &  \citet{Bennett2016}  \\
                     &  OGLE-2013-BLG-0341   &  \citet{Gould2014}    \\
                     &  OGLE-2016-BLG-0613   &  \citet{Han2017}      \\
                     &  OGLE-2018-BLG-1700   &  \citet{Han2020b}     \\ 
                     &  OGLE-2019-BLG-0304   &  \citet{Han2020g}     \\
\hline
3L1S                 &  OGLE-2006-BLG-109    &  \citet{Gaudi2008, Bennett2010}  \\
(two-planet system)  &  OGLE-2012-BLG-0026   &  \citet{Han2013}      \\
                     &  OGLE-2018-BLG-1011   &  \citet{Han2019}      \\
\hline
1L3S                 &  OGLE-2015-BLG-1459   &  \citet{Hwang2018}    \\
\hline
2L2S                 &  MOA-2010-BLG-117     &  \citet{Bennett2018}  \\
                     &  OGLE-2016-BLG-1003   &  \citet{Jung2017}     \\
\hline
3L2S                 &  KMT-2019-BLG-1715    &  \citet{Han2020e}     \\   
\hline
3L1S or 2L2S         &  OGLE-2014-BLG-1722   &  \citet{Suzuki2018}   \\
                     &  OGLE-2018-BLG-0532   &  \citet{Ryu2020}      \\
                     &  KMT-2019-BLG-1953    &  \citet{Han2020a}     \\
\hline
\end{tabular}
\end{table*}

\section{Introduction}\label{sec:one}

Microlensing light curves can exhibit deviations from the smooth and symmetric form of a
single-lens single-source (1L1S) event. The most common causes for these anomalies are the binary
nature of the lens, 2L1S event \citep{Mao1991}, and the source, 1L2S event \citep{Griest1992}. 
Detections of such three-object (2L+1S or 1L+2S) events by the first-generation lensing
experiments, e.g., MACHO LMC~1 \citep{Dominik1994} and OGLE~7 \citep{Udalski1994}
2L1S events and MACHO LMC 96-2 \citep{Becker1997} 1L2S event, prompted theoretical studies
on the lensing behavior under these lens system configurations. This led to the establishment of
observational strategies for efficient detections of microlensing anomalies, e.g., survey+follow-up
mode observations for intensive coverage of short-lasting anomalies \citep{Gould1992},
and the development of methodologies for efficient analyses of anomalous lensing events, e.g.,
ray-shooting method \citep{Bond2002, Dong2009, Bennett2010} and contour integration algorithm 
\citep{Gould1997, Bozza2018}.  Thanks to the accomplishments on both observational and theoretical 
sides, more than a hundred anomalous events are currently being detected each year, and they are 
promptly analyzed almost in real time with the progress of events \citep{Ryu2010, Bozza2012}

With the great increase of the event detection rate together with the dense coverage of lensing 
light curves by high-cadence lensing surveys, OGLE-IV \citep{Udalski2015}, MOA \citep{Bond2001}, 
and KMTNet \citep{Kim2016}, one is occasionally confronted with events for which observed lensing 
light curves cannot be explained by interpretations with three objects.  At the time of writing 
this article, there exist thirteen confirmed cases of events, for which at least four objects 
(lenses plus sources) are required to interpret observed light curves. These events are listed in 
Table~\ref{table:one}.  Nine of these are 3L1S events, in which the lensing system is composed of 
three lens masses and a single source star. Among them, the lenses of six events (OGLE-2008-BLG-092, 
OGLE-2007-BLG-349, OGLE-2013-BLG-0341, OGLE-2016-BLG-0613, OGLE-2018-BLG-1700, and OGLE-2019-BLG-0304) 
are planets in binaries, and the lenses of the other three events (OGLE-2006-BLG-109, OGLE-2012-BLG-0026, 
OGLE-2018-BLG-1011) are systems containing two planets.  The lensing event OGLE-2015-BLG-1459 was 
identified as a 1L3S event, in which a single-lens mass was involved with three source stars.  The 
events MOA-2010-BLG-117 and OGLE-2016-BLG-1003 were very rare cases, in which both the lens and source 
are binaries.  Interpretation of the event KMT-2019-BLG-1715 is even more complex and requires five 
objects, in which the lens is composed of three masses (a planet plus two stars) and the source 
consists of two stars, that is, 3L2S event.  Besides these events, there exist three additional 
events (OGLE-2014-BLG-1722, OGLE-2018-BLG-0532, and KMT-2019-BLG-1953), in which four-object modeling 
is required to explain the observed light curves, but unique solutions cannot be firmly specified due 
to either degeneracies among different interpretations or not enough coverage of signals.  Accumulation 
of knowledge from modeling these multi-body events is important for future interpretations of lensing 
light curves with complex anomalous features.

In this paper, we present the analysis of the lensing event KMT-2019-BLG-0797.  The light curve 
of the event exhibits two anomalous features, which cannot be explained by a usual 2L1S or 1L2S model.  
We test various four-object models, in which an extra lens or source are considered in the interpretation 
of the event.

The anomalous nature of KMT-2019-BLG-0797 was found from a project conducted to reanalyze previous 
KMTNet events detected in and before the 2019 season. In the first part of this project, \citet{Han2020d} 
investigated events involved with faint source stars and found four planetary events (KMT-2016-BLG-2364, 
KMT-2016-BLG-2397, OGLE-2017-BLG-0604, and OGLE-2017-BLG-1375), for which no detailed investigation had 
been conducted.  The second part of the project was focused on high-magnification events, aiming to find 
subtle planetary signals, and this led to the discoveries of two planetary systems KMT-2018-BLG-0748L 
\citep{Han2020c} and KMT-2018-BLG-1025L \citep{Han2020h}. The event KMT-2019-BLG-0797 was closely examined 
as a part of the project investigating high-magnification events involved with faint source stars.

For the presentation of the work, we organize the paper as follows. In Sect.~\ref{sec:two}, we mention the 
data of the lensing event analyzed in this work and describe observations conducted to acquire the data.  
In Sect.~\ref{sec:three}, we depict the anomaly that appeared on the light curve and describe various 
tests conducted to interpret the anomaly.  We estimate the angular Einstein radius of the lensing event in 
Sect.~\ref{sec:four}, and estimate the physical parameters of the lens and source in Sect.~\ref{sec:five}. 
We summarize the results and conclude in Sect.~\ref{sec:six}.

\section{Observation and data}\label{sec:two}

The lensing event KMT-2019-BLG-0797 occurred on a source lying toward the Galactic bulge.
The equatorial and galactic coordinates of the source are $({\rm RA}, {\rm DEC})_{\rm J2000}=
(17:53:48.09, -31:58:36.70)$ and $(l, b)=(-1^\circ\hskip-2pt .706, -3^\circ\hskip-2pt .073)$, 
respectively. The brightness of the source had remained constant with an apparent baseline 
brightness of $I_{\rm base}\sim 20.1$, as measured on the KMTNet scale, before the lensing 
magnification.  The lensing-induced magnification of the source flux lasted about 10 days as 
measured by the duration beyond the photometric scatter.

\begin{figure}
\includegraphics[width=\columnwidth]{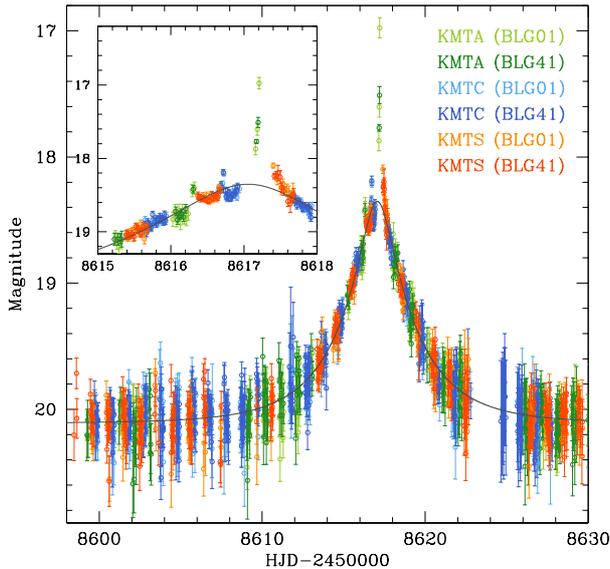}
\caption{
Light curve of KMT-2019-BLG-0797. The curve drawn over the data points is a 1L1S model. The inset 
shows the region around the peak.  The colors of the data points match those of the telescopes, 
marked in the legend, used to acquire the data.
}
\label{fig:one}
\end{figure}

The event was detected on 2019 May 13 (${\rm HJD}^\prime\equiv {\rm HJD}-2450000\sim 8616.6$) 
by the Alert Finder System \citep{Kim2018} of the KMTNet survey. The survey uses three identical 
wide-field telescopes that are globally located in three continents. The locations of the individual 
telescopes are the Siding Spring Observatory (KMTA) in Australia, the Cerro Tololo Inter-American 
Observatory (KMTC) in South America, and the South African Astronomical Observatory (KMTS) in Africa. 
Each KMTNet telescope has a 1.6m aperture and is equipped with a camera yielding $2^\circ \times 2^\circ$ 
field of view. Images of the source were mainly acquired in the $I$ band, and about one tenth of images 
were obtained in the $V$ band for the source color measurement.  We will describe the detailed procedure 
of determining the source color in Sect.~\ref{sec:four}.

Figure~\ref{fig:one} shows the lensing light curve of KMT-2019-BLG-0797. The curve plotted over 
the data points is a 1L1S model with $(t_0, u_0, \te)\sim (8617.06, 0.20, 4.6~{\rm days})$, where 
the individual lensing parameters indicate the peak time, impact parameter of the lens-source 
approach (normalized to the angular Einstein radius $\thetae$), and event timescale, respectively. 
The inset shows the zoomed-in view of the peak region, around which the data exhibit an anomaly 
relative to the 1L1S model.  The date of the event alert approximately corresponds to the peak of 
the lensing magnification. The anomaly was already in progress at the time of the event alert, but 
it was not noticed due to the subtlety of the anomaly together with the considerable photometric 
uncertainties of data caused by the faintness of the source. As a result, little attention was paid 
to the event when the event was found, and thus no alert for follow-up observations was issued.  
Nevertheless, the peak region of the light curve was densely and continuously covered, because the 
source was located in the two overlapping KMTNet fields of BLG01 and BLG41, toward which observations 
were conducted most frequently.  The observational cadence for each field was 30~min, and thus the 
event was covered with a combined cadence of 15~min.

Photometry of the event was conducted utilizing the KMTNet pipeline \citep{Albrow2009}, which 
is a customized version of pySIS code developed on the basis of the difference imaging method
\citep{Tomaney1996, Alard1998}. Additional photometry is conducted for a subset of KMTC $I$- and 
$V$-band data using the pyDIA software \citep{Albrow2017} to measure the color of the source and 
to construct a color-magnitude diagram (CMD) of ambient stars around the source. We readjust error 
bars of data estimated from the pipeline following the standard routine described in \citet{Yee2012}.

\begin{figure}
\includegraphics[width=\columnwidth]{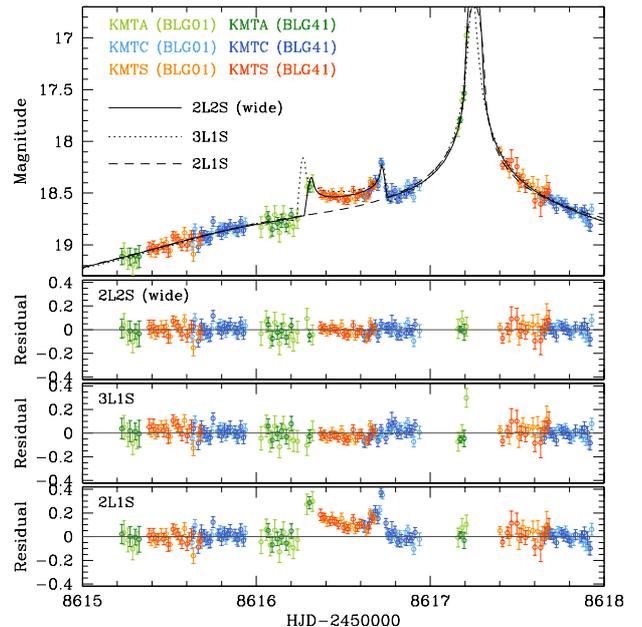}
\caption{
Zoomed-in view around the peak of the light curve. Drawn over the data points are the 
2L2S (wide), 3L1S, and 2L1S models. 
The lower three panels show the residuals from the individual models.
}
\label{fig:two}
\end{figure}

\section{Anomaly}\label{sec:three}

Figure~\ref{fig:two} shows the detailed pattern of the anomaly in the peak region. It shows that 
the anomaly consists of two distinctive features. The first is the caustic-crossing feature, 
which is composed of two spikes at ${\rm HJD}^\prime \sim 8616.3$ and 8616.7 and the U-shape trough 
between the spikes.  The other feature is the strong positive deviation from the 1L1S model centered 
at ${\rm HJD}^\prime \sim 8617.2$.  The KMTA data partially cover the rising side of the second 
feature, but the falling side was not covered until almost the end of the anomaly.

\subsection{2L1S model}\label{sec:three-one}

Considering that both anomalous features are likely to be involved with source star's crossing 
over or approaching a caustic, we first model the observed light curve under the assumption that 
the lens is a binary. In addition to the 1L1S lensing parameters, a 2L1S modeling requires one to 
include the three additional parameters $(s, q, \alpha)$, which represent the projected separation 
(normalized to $\thetae$), mass ratio between the binary lens components, $M_1$ and $M_2$, and the 
angle between the source trajectory and the $M_1$--$M_2$ axis (source trajectory angle), respectively.  
In the 2L1S modeling, we conduct thorough grid searches for the binary parameters $(s, q)$ with 
multiple starting points of $\alpha$ evenly distributed in the range of $0\leq \alpha\leq 2\pi$.  
For the computations of finite-source magnifications, we use the ray-shooting method described in 
\citet{Dong2009}.  From this investigation, we find that the 2L1S modeling does not yield a plausible 
model explaining the observed anomalous
features.

We then conduct another 2L1S modeling, this time, by excluding the data around one of the two anomaly 
features. The modeling conducted by removing the data around the second anomalous feature does not 
yield a reasonable solution either.\footnote{ As we will show in the following subsection, the first 
anomalous feature is produced by an extra source. This source comprises a very minor fraction of the 
primary source, and the most flux comes from the primary source.  As a result, the contribution of 
the second source to the lensing light curve is confined only to the time of the first anomaly, and 
thus the 2L1S modeling conducted excluding the second anomalous feature does not yield a model 
describing the overall light curve.} However, the modeling with the exclusion of the first anomalous 
feature yields a solution that well describes the second anomalous feature.

\begin{figure}
\includegraphics[width=\columnwidth]{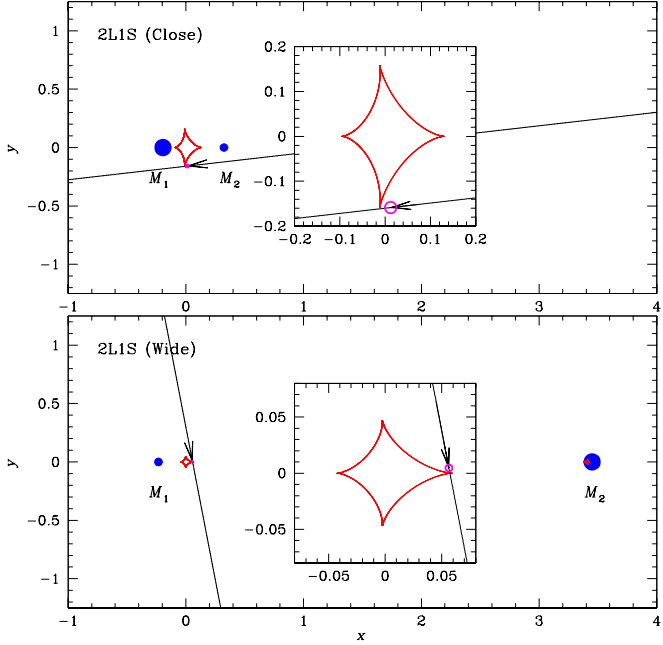}
\caption{
Lens system configurations of the 2L1S solutions. The upper and lower panels are for the close 
and wide solutions, respectively. The inset in each panel shows the enlarged view around the 
caustic located close to the source trajectory (line with an arrow). The blue dots marked by 
$M_1$ and $M_2$ denote the positions of the binary lens components, and the bigger dot represents 
the heavier lens mass.  Lengths are scaled to the Einstein radius.  The small magenta circle on the 
source trajectory represents the source size scaled to the Einstein radius: 
$\rho = \theta_*/\theta_{\rm E}$.
}
\label{fig:three}
\end{figure}

\begin{table}[htb]
\small
\caption{Best-fit parameters of 2L1S solutions\label{table:two}}
\begin{tabular*}{\columnwidth}{@{\extracolsep{\fill}}lll}
\hline\hline
\multicolumn{1}{c}{Parameter}                      &
\multicolumn{1}{c}{Close}                          &
\multicolumn{1}{c}{Wide}                           \\
\hline
$t_0$ (${\rm HJD}^\prime$)          &  $8617.114 \pm 0.007$   &  $8617.103 \pm 0.007                 $   \\
$u_0$ / $u_0^\prime$                &  $0.159 \pm 0.005   $   &  $0.059 \pm 0.004$ / $0.154 \pm 0.011$   \\
$\te$ / $t_{\rm E}^\prime$ (days)   &  $4.64 \pm 0.09     $   &  $12.56 \pm 0.66$ / $4.83 \pm 0.26   $   \\
$s$                                 &  $0.519 \pm 0.008   $   &  $3.628 \pm 0.051                    $   \\
$q$                                 &  $0.599 \pm 0.068   $   &  $5.753 \pm 0.705                    $   \\ 
$\alpha$ (rad)                      &  $6.168 \pm 0.013   $   &  $4.524 \pm 0.010                    $   \\
$\rho$ / $\rho^\prime$ ($10^{-3}$)  &  $12.72 \pm 1.73    $   &  $3.00 \pm 0.69$ / $7.81 \pm 1.82    $   \\
\hline
\end{tabular*}
\tablefoot{ ${\rm HJD}^\prime\equiv {\rm HJD}-2450000$.  
The parameters $(u_0^\prime, t_{\rm E}^\prime, \rho^\prime)$ of the wide solution are the values scaled to the
angular Einstein radius corresponding to $M_1$, i.e., $\theta_{\rm E}^\prime = \thetae/(1+q)^{1/2}$.
}
\end{table}

The model curve of the 2L1S solution obtained by excluding the first anomalous feature is plotted over 
the data points in Figure~\ref{fig:two}, and the lensing parameters of the solution are listed in 
Table~\ref{table:two}.  We find two sets of solutions, in which one has a binary separation smaller than 
unity $(s<1.0)$ and the other solution has a separation greater than unity ($s>1.0$). We refer to  the 
two solutions as the ``close'' and ``wide'' solutions, respectively. The binary parameters of the close 
and wide solutions are $(s, q, \alpha)_{\rm close}\sim (0.52, 0.60, -6.6^\circ)$ and $(s, q, \alpha)_{\rm wide} 
\sim (3.62, 5.75, -100.8^\circ)$, respectively.  We note that $M_1$ denotes the lens component located 
closer to the source trajectory, not the heavier mass component, and thus the mass ratio $q$ of the wide 
solution is greater than unity.  For the wide solution, we present additional parameters $(u_0^\prime, 
t_{\rm E}^\prime, \rho^\prime)$, which represent the values scaled to the angular Einstein radius 
corresponding to $M_1$, $\theta_{\rm E}^\prime=\thetae/(1+q)^{1/2}$, and thus $u_0^\prime=u_0(1+q)^{1/2}$, 
$t_{\rm E}^\prime =t_{\rm E}/(1+q)^{1/2}$, and $\rho^\prime =\rho (1+q)^{1/2}$. It is found that the 
parameters $(u_0^\prime, t_{\rm E}^\prime, \rho^\prime)$ of the wide solution are similar to the 
corresponding parameters of the close solution.

\begin{figure}
\includegraphics[width=\columnwidth]{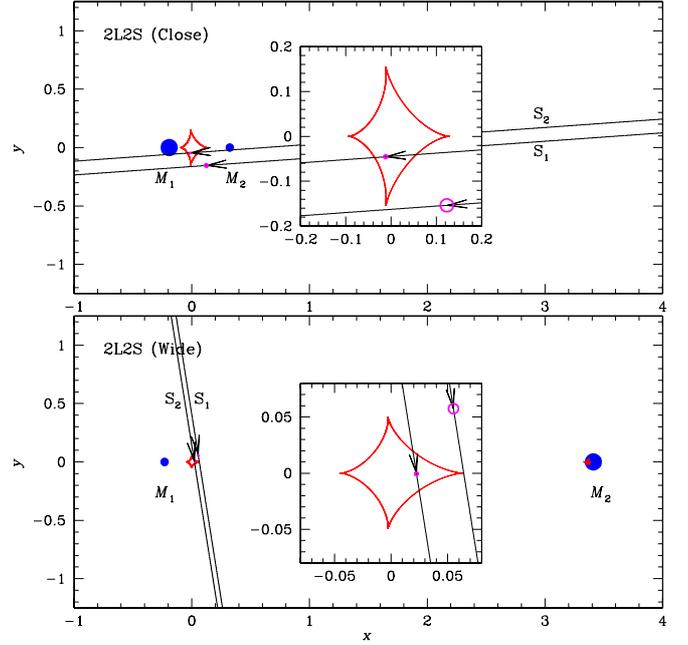}
\caption{
Lens system configuration of the 2L2S model. Notations are same as those in Fig.~\ref{fig:three}
except that there is an additional source trajectory of the second source, $S_2$.
}
\label{fig:four}
\end{figure}

\begin{table}[t]
\small
\caption{Best-fit parameters of 2L2S solutions\label{table:three}}
\begin{tabular*}{\columnwidth}{@{\extracolsep{\fill}}lll}
\hline\hline
\multicolumn{1}{c}{Parameter}                      &
\multicolumn{1}{c}{Close}                          &
\multicolumn{1}{c}{Wide}                           \\
\hline
$\chi^2$                                  &  $1907.2             $   &  $1901.7$                               \\
$t_{0,1}$ (${\rm HJD}^\prime$)            &  $8617.145 \pm 0.008 $   &  $8617.132 \pm 0.006 $                  \\
$u_{0,1}$ / $u_{0,1}^\prime$              &  $0.162 \pm 0.006    $   &  $0.063 \pm 0.003$ / $0.158 \pm 0.005$  \\
$t_{0,2}$ (${\rm HJD}^\prime$)            &  $8616.549 \pm 0.007 $   &  $8616.500 \pm 0.005$                   \\
$u_{0,2}$ / $u_{0,2}^\prime$              &  $0.045 \pm 0.002    $   &  $0.022 \pm 0.001$ / $0.055 \pm 0.003$  \\ 
$\te$ / $t_{\rm E}^\prime$ (days)         &  $4.68 \pm 0.08      $   &  $12.08 \pm 0.35$ / $4.82 \pm 0.14$     \\
$s$                                       &  $0.515 \pm 0.008    $   &  $3.641 \pm 0.042$                      \\
$q$                                       &  $0.595 \pm 0.052    $   &  $5.259 \pm 0.387$                      \\
$\alpha$ (rad)                            &  $6.211 \pm 0.011    $   &  $4.556 \pm 0.009$                      \\
$\rho_1$ / $\rho_1^\prime$ ($10^{-3}$)    &  $14.24 \pm 2.33     $   &  $4.38 \pm 0.45$ / $10.96 \pm 1.13$     \\
$\rho_2$ / $\rho_2^\prime$ ($10^{-3}$)    &  $3.99 \pm 0.50      $   &  $1.45 \pm 0.19$ / $3.63 \pm 0.48 $     \\ 
$q_{F,I}$                                 &  $0.040 \pm 0.003    $   &  $0.043 \pm 0.003$                      \\
\hline
\end{tabular*}
\end{table}

Figure~\ref{fig:three} shows the configuration of the lens system corresponding to the 2L1S solutions. The 
upper and lower panels are the configurations of the close and wide solutions, respectively.  According 
to these solutions, the second feature of the anomaly is produced by the source crossing over the tip of 
the four-cusp caustic induced by a binary. We note that the binary separation, $s\sim 0.52$ for the close 
solution and $s\sim 2.6$ for the wide solution, of the lens is neither much smaller nor much greater than 
unity, and thus the lens system is not in the regime of the Chang-Refsdal (C-R) lensing, which refers to 
the gravitational lensing of a point mass perturbed by a constant external shear \citep{Chang1979, Chang1984}. 
As a result, the caustic slightly deviates from the symmetric astroid shape of the C-R lensing caustic. The 
source crosses the caustic tip located on the long side of the caustic

\subsection{2L2S model}\label{sec:three-three}

We conduct another modeling under the interpretation that both the lens and source are binaries
(2L2S model). The lensing magnification of a 2L2S event is the superposition of those involved
with the individual source stars, $S_1$ and $S_2$, that is,
\begin{equation}
A={A_1 F_{S_1}+A_2 F_{S_2} \over F_{S_1}+F_{S_2}} =
  {A_1 + A_2  q_F \over 1 + q_F  } .
\label{eq1}
\end{equation}
Here $(F_{S_1}, F_{S_2})$ and $(A_1, A_2)$ denote the baseline flux values and the magnifications 
associated with the individual source stars, respectively, and $q_F=F_{S_2}/F_{S_1}$ represents the 
flux ratio between the source stars. The consideration of an extra source requires one to include 
additional parameters in modeling. These parameters are $(t_{0,2}, u_{0,2}, \rho_2, q_F)$, which 
represent the time of the closest approach of $S_2$ to the lens, and the lens-source separation 
at that time, the normalized radius of $S_2$, and the flux ratio between $S_1$ and $S_2$, respectively 
\citep{Hwang2013}. As initial values of the parameters related to the $S_1$, ($t_{0,1}, u_{0,1}, 
\te, s, q, \alpha, \rho_1)$, we use the values obtained from the 2L1S modeling. We set the initial 
values of $(t_{0,2}, u_{0,2}, \rho_2, q_F)$ considering the time and strength of the first anomaly 
feature.

\begin{figure}
\includegraphics[width=\columnwidth]{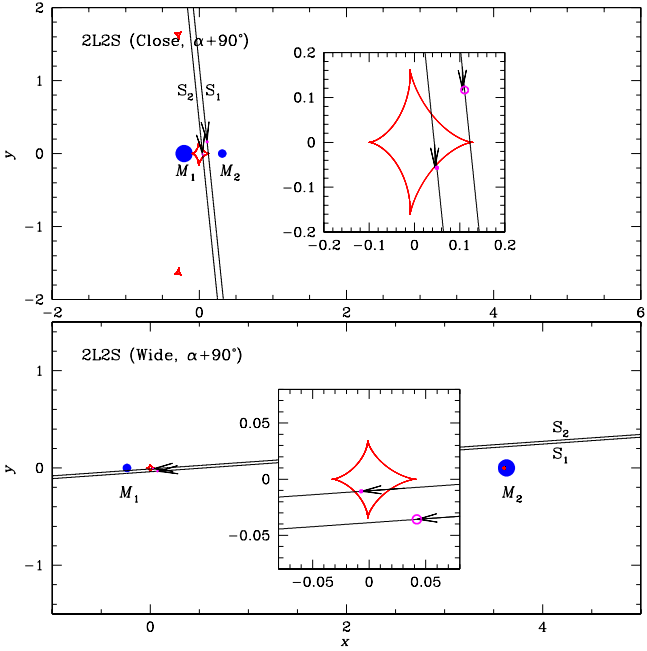}
\caption{
Lens system configuration of the models obtained by confining the source trajectory angle around 
$\sim \alpha + 90^\circ$ from the best-fit solutions with $\alpha$.
}
\label{fig:five}
\end{figure}

It is found that the 2L2S modeling yields solutions that well describe the observed data including 
both anomalous features. We find two sets of solutions resulting from the close--wide degeneracy, and 
the best-fit lensing parameters of the individual solutions are listed in Table~\ref{table:three}.  
It is found that the wide solution yields a slightly better fit to the data than the close 
solution, especially in the region around the second anomalous feature.  However, the $\chi^2$ difference 
between the two solutions is merely $\Delta\chi^2=5.5$, and thus we consider the close solution as 
a viable model.  The model curves of the wide solution and the residual from the model are shown in 
Figure~\ref{fig:two}.

\begin{figure}
\includegraphics[width=\columnwidth]{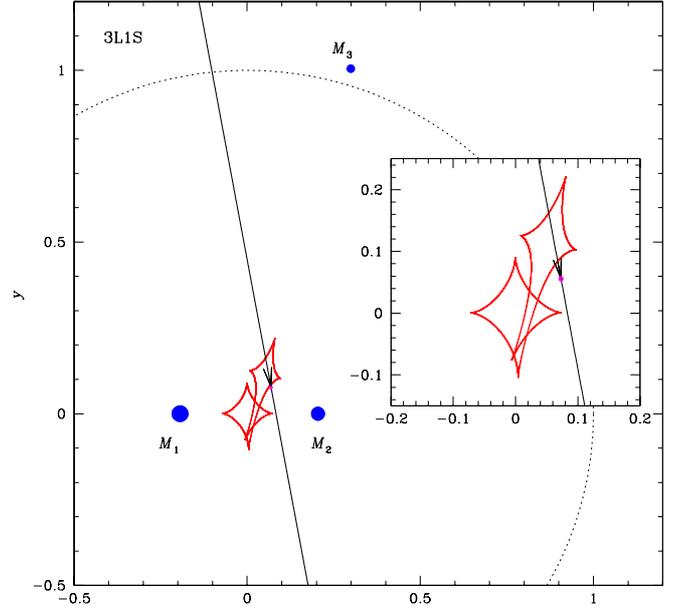}
\caption{
Lens system configuration of the 3L1S model.  We note that there are three lens components, 
marked by $M_1$, $M_2$, and $M_3$.  The dotted circle with a radius unity and centered at the 
$M_1$--$M_2$ barycenter represents the Einstein ring.
}
\label{fig:six}
\end{figure}

\begin{table}[htb]
\small
\caption{Best-fit parameters of 3L1S solution\label{table:four}}
\begin{tabular*}{\columnwidth}{@{\extracolsep{\fill}}ll}
\hline\hline
\multicolumn{1}{c}{Parameter}                      &
\multicolumn{1}{c}{Value}                           \\
\hline
$\chi^2$                      &   2028.6                     \\
$t_{0}$ (${\rm HJD}^\prime$)  &  $8617.159 \pm  0.006  $     \\
$u_{0}$                       &  $   0.082 \pm  0.004  $     \\
$\te$  (days)                 &  $   6.03  \pm  0.22   $     \\
$s_2$                         &  $   0.398 \pm  0.008  $     \\
$q_2$                         &  $   0.947 \pm  0.064  $     \\
$\alpha$ (rad)                &  $   4.529 \pm  0.011  $     \\
$s_3$                         &  $   1.119 \pm  0.006  $     \\
$q_3$ ($10^{-3}$)             &  $   2.75  \pm  0.28   $     \\
$\psi$ (rad)                  &  $   1.115 \pm  0.011  $     \\
$\rho$ ($10^{-3}$)            &  $   2.70  \pm  0.37   $     \\
\hline
\end{tabular*}
\end{table}

In Figure~\ref{fig:four}, we present the lens system configurations corresponding to the close (upper 
panel) and wide (lower panel) 2L2S solutions. From the comparison of the lensing parameters with those of 
the 2L1S solutions, presented in Table~\ref{table:two}, it is found that the parameters related to $S_1$ 
$(t_{0,1}, u_{0,1}, \te, s, q, \alpha)$ for the 2L1S and 2L2S solutions are similar to each other. The 
main difference between the two solutions is the presence of an additional source $S_2$ that approaches 
$M_1$ closer than $S_1$ does. The second source passes over the caustic producing a caustic-crossing 
feature in the light curve, and this explains the first anomalous feature that could not be explained by 
the 2L1S model. According to the 2L2S solutions, the companion source comprises about 4\% of the $I$-band 
flux from the primary source, that is, $q_{F,I}\sim 0.04$.

We note that the source trajectory of the event is well constrained despite the approximately
symmetric shape of the caustic. For a binary lens in a C-R lensing regime, the induced caustic has
a symmetric astroid shape with four cusps. In this case, the lensing light curves resulting from the 
source trajectory angles $\alpha$ and $\alpha\pm 90^\circ$ have a similar shape. See Figure~4 of 
\citet{Hwang2010} for the illustration of this degeneracy.  We find that KMT-2019-BLG-0797 is not 
subject to this degeneracy because the source trajectory of the solution with $\alpha\pm 90^\circ$ 
is approximately aligned with the line connecting the central and peripheral caustics of the binary 
lens.  To demonstrate this, in Figure~\ref{fig:five}, we plot the lens system configurations of the 
solutions obtained by confining the source trajectory angle around $\sim \alpha\pm 90^\circ$ from the 
best-fit solution. It shows that these solutions result in source trajectories passing close to the 
peripheral caustic lying away from the central caustic.  The source approach to the peripheral caustic 
results in an additional bump in the lensing light curve before the main peak.  To avoid such a bump, 
the source trajectory angle of these solutions has less freedom in $\alpha$, and this leads to a 
worse fit than the solutions presented in Table~\ref{table:three}.

\subsection{3L1S model}\label{sec:three-two}

We also test a model, in which the lens has three components (3L1S model).  We test this model 
because the anomalous features appear around the peak of the light curve, and thus a third 
mass, if it exists, may induce an additional caustic and explain the first anomalous feature that 
is not explained by a 2L1S model, for example, OGLE-2016-BLG-0613 \citep{Han2017}, OGLE-2018-BLG-1700 
\citep{Han2020f}, OGLE-2019-BLG-0304 \citep{Han2020g}.  In the 3L1S modeling, we conduct a thorough 
grid search for the parameters describing the third lens mass, $M_3$. These parameters include 
$(s_3, q_3, \psi)$, which denote the normalized $M_1$--$M_3$ separation, mass ratio $q_3=M_3/M_1$, 
and the position angle of $M_3$ as measured from the $M_1$--$M_2$ axis and with a center at $M_1$.

\begin{figure}
\includegraphics[width=\columnwidth]{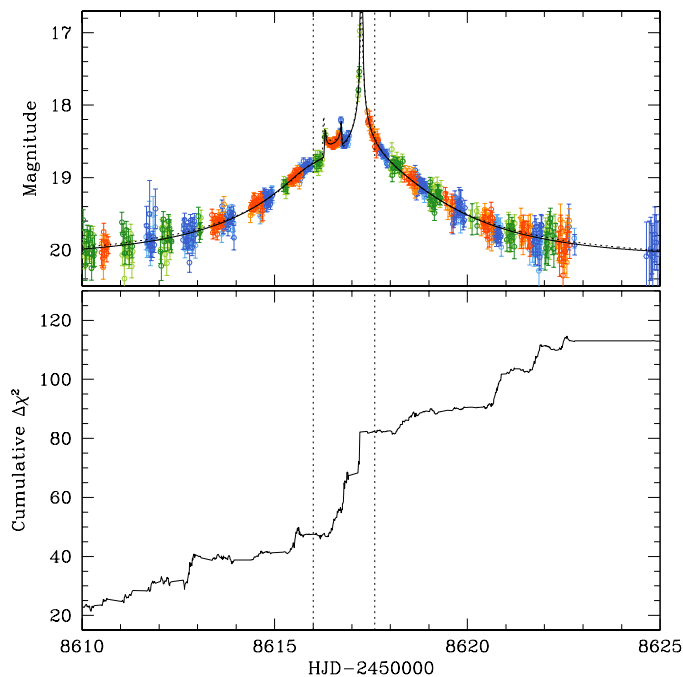}
\caption{
Cumulative distribution of $\chi^2$ difference between the 3L1S and 2L2S models, that is, 
$\Delta\chi^2=\chi^2_{\rm 3L1S}-\chi^2_{\rm 2L2S}$.  The two dotted vertical lines are drawn to 
indicate the region of the anomalies.
}
\label{fig:seven}
\end{figure}

The 3L1S modeling also yields a solution that appears to depict both anomalous features.  In 
Figure~\ref{fig:two}, we present the model curve and residual in the peak region of the light 
curve.  The best-fit lensing parameters of the model are listed in Table~\ref{table:four}, and 
the corresponding lens system configuration is shown in Figure~\ref{fig:six}.  According to this 
solution, the first anomalous feature is produced by the crossing of the source over an additional 
caustic  induced by a low-mass third lens component.  The estimated mass ratio of the third mass 
to the primary is $q_3=M_3/M_1=(2.75\pm 0.28) \times 10^{-3}$, indicating that $M_3$ is a planetary 
mass object. The planet is located close to the Einstein ring corresponding to $M_1 + M_2$ with a 
position angle of $\psi\sim 64^\circ$.  Due to the proximity of $s_3$ to unity, the planet induces 
a single large resonant caustic, and the source passes through the planet-induced caustic, producing 
the first anomalous feature, before it approaches the cusp of the binary-induced caustic, producing 
the second anomalous
feature.

Although the 3L1S model seemingly describes the anomalous features, it is found that the model fit 
is substantially worse than the 2L2S model.  The difference in the fits between the two models  
as measured by $\chi^2$ difference is $\Delta\chi^2=\chi^2_{\rm 3L1S}-\chi^2_{\rm 2L2S}=126.9$, 
indicating that the 2L2S model is strongly preferred over the 3L1S model.  To show the difference 
in the fits, we present the cumulative distribution of $\Delta\chi^2$ between the two models in 
Figure~\ref{fig:seven}.  The distribution shows that the 2L2S model provides a better fit than the 
3L1S model not only in the region around the peak but also throughout the lightcurve during the 
lensing magnification.

\section{Angular Einstein radius}\label{sec:four}

\begin{figure}
\includegraphics[width=\columnwidth]{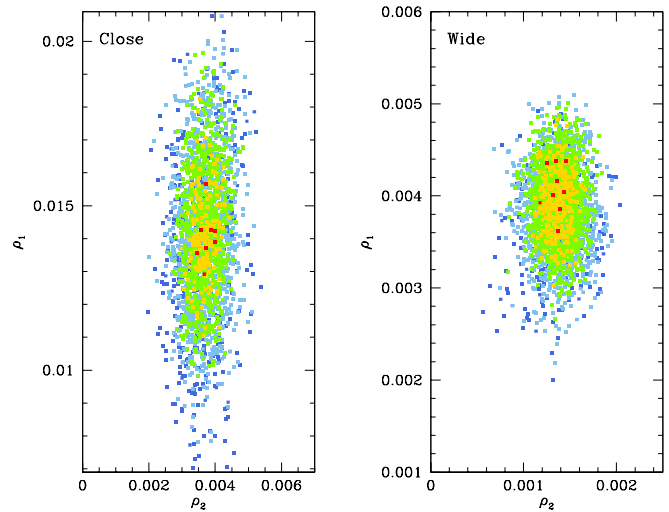}
\caption{
Scatter plots in the MCMC chain on the $\rho_1$--$\rho_2$ parameter plane for the close (left panel) 
and wide (right panel) solutions. The color coding is set to indicate points within 1$\sigma$ (red), 
2$\sigma$ (yellow), 3$\sigma$ (green), 4$\sigma$ (cyan), and 5$\sigma$ (blue).
}
\label{fig:eight}
\end{figure}

In this section, we estimate the angular Einstein radius $\thetae$. In order to estimate $\thetae$, 
it is required to measure the normalized source radius, that is related to the angular Einstein 
radius by
\begin{equation}
\thetae = {\theta_* \over \rho}.
\label{eq2}
\end{equation}
Here $\theta_*$ represents the angular source radius.  We find that the normalized source radii of 
both $S_1$ and $S_2$ are constrained, although the uncertainties are considerable due to a partial
coverage of the caustic crossings.  This can be seen in Figure~\ref{fig:eight}, in which we present 
scatter plots of the points in the MCMC chain on the $\rho_1$--$\rho_2$ parameter plane for the close 
(left panel) and wide (right panel) solutions.  We note that the uncertainty of $\rho_1$ is greater 
than the uncertainty of $\rho_2$, because only the rising part of the second 
anomalous feature was covered by the data.

Another requirement for the $\thetae$ measurement is estimating the angular source radius $\theta_*$. 
We estimate $\theta_*$ from the color and brightness of the source. In order to estimate calibrated 
color and brightness from the instrumental values, we use the method of \citet{Yoo2004}. In this method, 
the centroid of red giant clump (RGC), with its known de-reddened color and magnitude, in the CMD serves 
as a reference for the color and magnitude calibration.  

\begin{figure}
\includegraphics[width=\columnwidth]{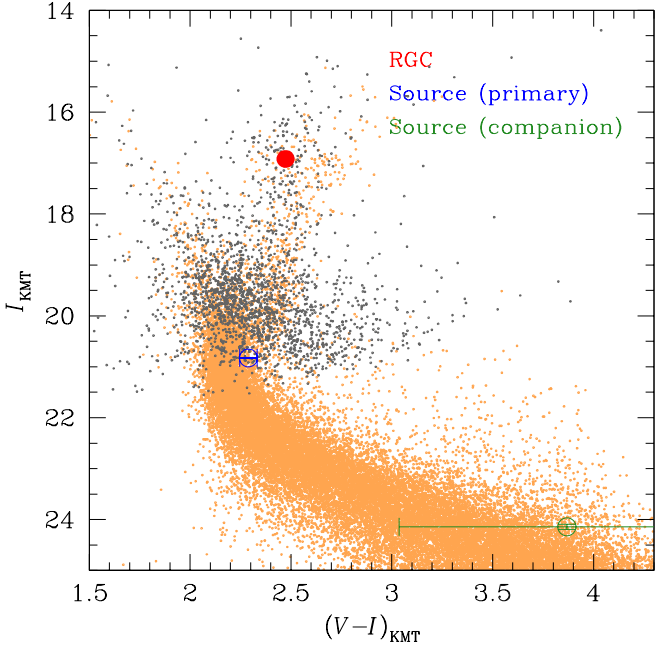}
\caption{
Locations of the primary and companion source stars with respect to the centroid of the red giant 
clump (RGC) in the instrumental color-magnitude diagram (CMD, grey dots).  The determinations of the 
source positions and the construction of the CMD are based on the pyDIA photometry of the KMTC data set.  
We also present the {\it Hubble Space Telescope} CMD \citep[][brown dots]{Holtzman1998} to show the 
source locations on the main-sequence branch. 
}
\label{fig:nine}
\end{figure}

\begin{table}[t]
\small
\caption{Source color and magnitude\label{table:five}}
\begin{tabular*}{\columnwidth}{@{\extracolsep{\fill}}ll}
\hline\hline
\multicolumn{1}{c}{Quantity}                           &
\multicolumn{1}{c}{Value}                                                 \\
\hline
$(V-I, I)_{\rm RGC}   $     &   $(2.473, 16.918)$                         \\
$(V-I, I)_{\rm RGC,0} $     &   $(1.060, 14.509)$                         \\
\hline
primary ($S_1$)             &                                             \\   
$(V-I, I)   $               &  $(2.289 \pm 0.043, 20.827 \pm 0.010)$      \\
$(V-I, I)_0 $               &  $(0.876 \pm 0.043, 18.418 \pm 0.010)$      \\
\hline
companion ($S_2$)           &                                              \\   
 $(V-I, I)  $               &   $(3.868 \pm 0.831, 24.143 \pm 0.053)$      \\
 $(V-I, I)_0$               &   $(2.455 \pm 0.831, 21.734 \pm 0.053)$      \\
\hline
\end{tabular*}
\end{table}

In the first step of the method, we estimate the combined (instrumental) flux from the source stars, 
$F_{S,p}=F_{S_1,p}+F_{S_2,p}$, and the companion/primary flux ratio, $q_{F,p}$, by fitting the KMTC 
photometry data set processed using the pyDIA code.  Here the subscript ``$p$'' denotes the passband 
of observation.  The measured values are 
$(F_{S,I}, F_{S,V})=(0.07751 \pm  0.00060,  0.00909 \pm 0.00032)$ and 
$(q_{F,I}, q_{F,V}) = (0.0471 \pm  0.0023, 0.0110 \pm 0.0084)$.  
We note that the value of $q_{F,I}$, which is derived from the pyDIA reduction, is slightly different 
(less than 1$\sigma$) from the value presented in Table~\ref{table:three}, which is derived from the 
pySIS reduction, because they come from different reductions of the data.  Then, the flux values from 
$S_1$ and $S_2$ are estimated by
\begin{equation}
F_{S_1,p}=\left( { 1\over 1+q_{F,p}}\right)F_{S,p};\qquad
F_{S_2,p}=\left( { q_{F,p}\over 1+q_{F,p}}\right)F_{S,p},
\label{eq3}
\end{equation}
respectively.  
From the measured flux values in the $I$ and $V$ bands, it is estimated that the colors and magnitudes 
of $S_1$ and $S_2$ are 
$(V-I,I)_{S_1}=(2.289 \pm 0.043, 20.827 \pm 0.010)$ and 
$(V-I,I)_{S_2}=(3.868 \pm 0.831, 24.143 \pm 0.053)$, respectively.  
In Figure~\ref{fig:nine}, we mark the locations of $S_1$ and $S_2$ with respect to the RGC centroid 
in the instrumental CMD constructed using the pyDIA photometry of the KMTC data (grey dots).

\begin{table}[t]
\small
\caption{Angular source radius, Einstein radius, and proper motion\label{table:six}}
\begin{tabular*}{\columnwidth}{@{\extracolsep{\fill}}lll}
\hline\hline
\multicolumn{1}{c}{Quantity}                   &
\multicolumn{1}{c}{Close}                      &
\multicolumn{1}{c}{Wide}                       \\
\hline
$\theta_*$ ($\mu$as)                       &   $0.803 \pm 0.066$    &  $\leftarrow$                          \\
$\thetae$ / $\theta^\prime_{\rm E}$ (mas)  &   $0.056 \pm 0.005$    &  $0.183 \pm 0.015$ / $0.073 \pm 0.006$ \\
$\mu$ (mas~yr$^{-1}$)                      &   $4.40 \pm 0.36  $    &  $5.54 \pm 0.46  $                     \\
\hline
\end{tabular*}
\tablefoot{The notation ``$\leftarrow$'' in the wide solution column implies that the value is 
same as the one in the left column. }
\end{table}

Although the $V$-band flux of $S_2$ has a relatively large {\it fractional} (i.e., magnitude) error, 
it is strongly constrained to be faint in an absolute sense. In particular, we find
\begin{equation}
q_{F,I} - q_{F,V} = 0.0381 \pm 0.0087,
\label{eq4}
\end{equation}
i.e., a $4.4\,\sigma$ difference. This serves as a second line of evidence that the 2L2S solution is
correct. If, for example, we had somehow missed a 3L1S solution, and the derived 2L2S solution
were merely mimicking it, then $q_{F,I} - q_{F,V}$ should be consistent with zero because there
is only one source with just one color. However, according to Equation~(\ref{eq4}) this possibility 
is excluded at $4.4\sigma$.

In the second step, we calibrate the color and magnitude.  With the measured instrumental color and 
brightness of the source, $(V-I, I)$, and the RGC centroid, $(V-I, I)_{\rm RGC}=(2.473, 16.918)$, together with the 
known de-reddened values of the RGC centroid, $(V-I, I)_{\rm RGC,0}=(1.060, 14.509)$  \citep{Bensby2013, 
Nataf2013}, the reddening and extinction corrected color and magnitude of the source are estimated by
\begin{equation}
(V-I, I)_0 = (V-I, I)_{RGC,0} + \Delta (V-I, I), 
\label{eq5}
\end{equation}
where $\Delta (V-I, I)$ denote the offsets in the color and brightness of the source from the 
RGC centroid measured on the instrumental CMD. This results in
\begin{equation}
(V-I, I)_0 =
\begin{cases}
(0.876 \pm 0.043, 18.418 \pm 0.010)  & \text{for $S_1$},\\
(2.455 \pm 0.831, 21.734 \pm 0.053)  & \text{for $S_2$}.  
\end{cases}
\label{eq6}
\end{equation} 
In Table~\ref{table:five}, we summarize the values of 
$(V-I, I)_{\rm RGC}$, $(V-I, I)_{\rm RGC,0}$, $(V-I, I)$, and $(V-I, I)_0$ 
for the primary and companion source stars.

According to the estimated color and magnitude, the fainter source, $S_2$, is clearly an early-to-mid 
M dwarf.  On the other hand, the brighter source, $S_1$, presents something of a puzzle because its 
best-fit position lies in a relatively unpopulated portion of the CMD (see Figure~\ref{fig:nine}). 
The most likely explanation is that its true position lies (1.5 -- 2.0)$\sigma$ blueward of the best-fit 
position. That is, it is most likely a very late G dwarf or a very early K dwarf. Because the position of 
$S_1$ is consistent with lying in the normal bulge population at the (1 -- 2)$\sigma$ level, we evaluate 
its angular radius using the measured values and normal error propagation.

In the third step, we estimate the angular source radius using the measured source color and brightness.  
For this, we first convert $V-I$ color into $V-K$ color using the color--color relation of \citet{Bessell1988}, 
and then estimate $\theta_*$ using the $(V-K)$-- $\theta_*$ relation of \citet{Kervella2004}. This process 
yields an angular source radius of
\begin{equation}
\theta_* =
0.803 \pm 0.066~\mu{\rm as}.
\label{eq7}
\end{equation}
We note that the source radius of $S_2$ is uncertain due to its large color uncertainty, and thus we use 
$\theta_*$ of $S_1$ for the $\thetae$ estimation, i.e., $\thetae=\theta_{*,1}/\rho_1$.  With the angular 
source radius, the angular Einstein radius is estimated using the relation in Equation~(\ref{eq2}), which 
yields 
\begin{equation}
\thetae =
\begin{cases}
0.056 \pm 0.005~{\rm mas}  & \text{(close)}, \\
0.073 \pm 0.006~{\rm mas}  & \text{(wide)}. 
\end{cases}
\label{eq8}
\end{equation}
Here the angular Einstein radius of the wide solution is scaled to $\theta_{\rm E}^\prime$. Together with 
the event timescale, the relative lens-source proper motion is estimated as
\begin{equation}
\mu = {\thetae \over \te} = 
\begin{cases}
4.40 \pm 0.36~{\rm mas}~{\rm yr}^{-1}  & \text{(close)}, \\
5.54 \pm 0.46~{\rm mas}~{\rm yr}^{-1}  & \text{(wide)}. 
\end{cases}
\label{eq9}
\end{equation}
In Table~\ref{table:six}, we summarize the values of $\theta_*$, $\thetae$, and $\mu$ corresponding 
to the close and wide solutions.

\begin{figure}
\includegraphics[width=\columnwidth]{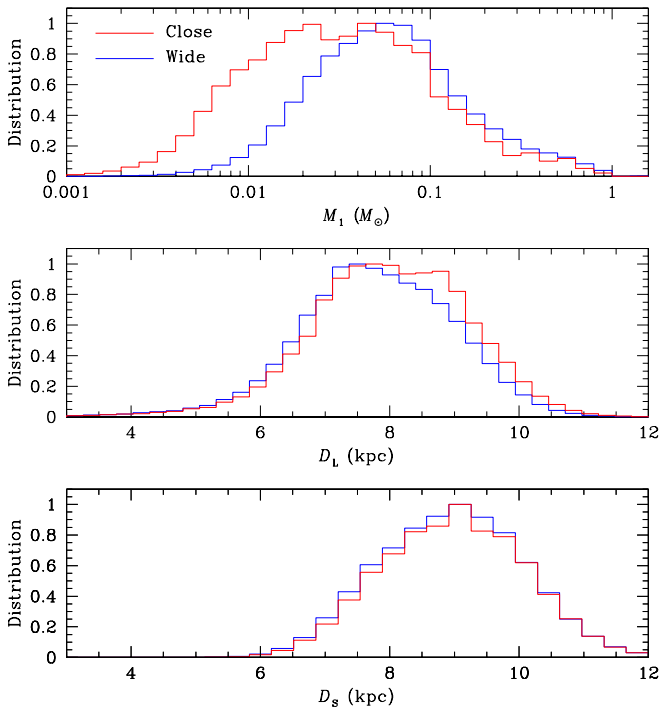}
\caption{
Bayesian posteriors of the primary lens mass ($M_1$), and distances to the lens ($D_{\rm L}$) and 
source ($D_{\rm S}$).  Red and blue curves are distributions obtained from the close and wide solutions, 
respectively.
}
\label{fig:ten}
\end{figure}

\section{Physical parameters of the lens and source}\label{sec:five}

For KMT-2019-BLG-0797, it is difficult to uniquely determine the physical lens parameters because
the microlens parallax cannot be measured due to the short timescale of the event, which is 
$\lesssim 5$~days. However, the lens parameters can still be constrained with the measured observables 
of the event timescale and angular Einstein radius, because these observables are related to the lens
parameters by
\begin{equation}
\te = {\thetae \over \mu },\ \ \ 
\thetae = (\kappa M \pi_{\rm rel})^{1/2}, \ \ \ 
\pi_{\rm rel}={\rm AU}\left( {1\over D_{\rm L}} - {1\over D_{\rm S}} \right). 
\label{eq10}
\end{equation}
Here $\kappa=4G/(c^2{\rm AU})$, $\pi_{\rm rel}$ denotes the relative lens-source parallax, $D_{\rm L}$ 
and $D_{\rm S}$ represent the distances to the lens and source, respectively.

\begin{table}[t]
\small
\caption{Physical lens parameters\label{table:seven}}
\begin{tabular*}{\columnwidth}{@{\extracolsep{\fill}}lcc}
\hline\hline
\multicolumn{1}{c}{Parameter}                      &
\multicolumn{1}{c}{Close}                          &
\multicolumn{1}{c}{Wide}                           \\
\hline
$M_1$ ($M_\odot$)         &  $0.034^{+0.076}_{-0.024} $   &   $0.064^{+0.107}_{-0.039} $ \\
$M_2$ ($M_\odot$)         &  $0.021^{+0.045}_{-0.014} $   &   $0.334^{+0.563}_{-0.204} $ \\
$D_{\rm L}$ (kpc)         &  $8.150^{+1.161}_{-1.161} $   &   $7.725^{+1.188}_{-1.134} $ \\
$D_{\rm S}$ (kpc)         &  $9.147^{+1.092}_{-1.158} $   &   $9.147^{+1.092}_{-1.158} $ \\ 
$a_{{\rm L},\perp}$ (AU)  &  $0.258^{+0.037}_{-0.037} $   &   $2.678^{+0.401}_{-0.383} $ \\
$a_{{\rm S},\perp}$ (AU)  &  $0.097^{+0.014}_{-0.014} $   &   $0.056^{+0.008}_{-0.008} $ \\
\hline
\end{tabular*}
\tablefoot{$M_1$ denotes the lens component located closer to the source trajectory,
not the heavier mass component.}
\end{table}

We estimate the physical lens parameters by conducting a Bayesian analysis. The analysis is done 
using the priors of the lens mass function and the physical and dynamical galactic models.  We 
adopt the mass function models of \citet{Zhang2020} and \citet{Gould2000}, in which the former mass 
function includes stellar and brown-dwarf lenses and the latter one accounts for stellar remnants. 
The locations and motions of lenses and source stars are assigned using the physical distribution 
model of \citet{Han2003} and the dynamical distribution model of \citet{Han1995}, respectively.  
Based on these models, we produce a large number ($4\times 10^7$) of artificial lensing events by 
conducting a Monte Carlo simulation, and then construct the distributions of the physical parameters.  
With the distributions, the representative values of the lens parameters are estimated as the median 
values of the distributions, and the uncertainties are estimated as the 16\% and 84\% ranges of the 
distributions.

Figure~\ref{fig:ten} shows the Bayesian posteriors of the primary lens mass and the distances 
to the lens and source.  In Table~\ref{table:seven}, we list the estimated parameters $M_1$, $M_2$, 
$D_{\rm L}$, $D_{\rm S}$, $a_{{\rm L},\perp}$, and $a_{{\rm S},\perp}$.  The last two parameters 
indicate the projected binary-lens and binary-source separations, that is,
\begin{equation}
a_{{\rm L},\perp}= sD_{\rm L}\thetae, \qquad 
a_{{\rm S},\perp}= \Delta u D_{\rm S}\thetae,
\label{eq11}
\end{equation}
where $\Delta u=\{ [(t_{0,1}-t_{0,2})/\te]^2+(u_{0,1}-u_{0,2})^2\}^{1/2}$ is the instantaneous 
separation between $S_1$ and $S_2$ at the time of the lensing magnification.

According to the close solution, the lens is composed of two brown dwarfs with masses $(M_1, M_2)\sim 
(0.034, 0.021)~M_\odot$ located in the bulge with a distance of $\dl= 8.2^{+1.2}_{-1.2}$~kpc.  According 
to the wide solution, on the other hand, the lens is composed of an object at the star/brown-dwarf 
boundary and an M dwarf with masses $(M_1, M_2)\sim (0.06, 0.33)~M_\odot$, and it is located at a 
distance of  $\dl=7.7^{+1.2}_{-1.1}$~kpc.  The masses of $M_1$ estimated from the close and wide 
solutions are similar to each other, although the wide solution prefers somewhat larger mass due to 
the larger value of the estimated $\thetae$.  On the other hand, the masses of $M_2$ estimated from 
the two degenerate solutions are widely different from each other.  This is because $M_1$ and $M_2$ 
have similar masses, with $q\sim 0.6$, according to the close solution, while $M_2$ according to the 
wide solution is much heavier than $M_1$, with $q\sim 5.3$.  For both the close and wide solutions, 
the source is located slightly behind the galactic center at a distance of $D_{\rm S}\sim 9.1$~kpc.  
This is because source stars located in the far side of the bulge have higher chances to be lensed 
than stars located in the front side.  The projected separations between the lens and source components 
are $(a_{{\rm L},\perp}, a_{{\rm S},\perp})\sim (0.26, 0.10)$~AU according to the close solution, and 
$\sim  (2.68, 0.06)$~AU according to the wide solution.  We note that the expected orbital period of 
the source, $P\gtrsim 10.5$~days for the close solution and $P\gtrsim 5$~days for the wide solution 
with the source mass of $M_S=M_{S_1}+M_{S_2}\sim 0.9~M_\odot + 0.3~M_\odot\sim 1.2~M_\odot$, is short, 
and thus the orbital motion of the source may affect the lensing light curve.  However, it is difficult 
to constrain the orbital motion first because the duration of the anomaly, which is $\sim 1$~day, is 
much shorter than the orbital period, and second because the photometric precision in the wings of 
the light curve is not good enough to capture the small modulations induced by the source orbital 
motion.

\section{Summary and conclusion}\label{sec:six}

We investigated the lensing event KMT-2019-BLG-0797, for which the light curve was found to be 
anomalous from the reexamination of events detected in and before the 2019 season.  For this event, 
it was found that a 2L1S model could not explain the anomaly.  From the tests with various models, 
it was found that the anomaly could be explained by introducing an extra source star to a 2L1S model.  
The event is the third case of a confirmed 2L2S event following on MOA-2010-BLG-117 
\citep{Bennett2018} and OGLE-2016-BLG-1003 \citep{Jung2017}.

The interpretation of the light curve was subject to a close--wide degeneracy.  According to the 
close solution, the lens is a binary consisting of two brown dwarfs with masses 
$(M_1, M_2)\sim (0.034, 0.021)~M_\odot$, and it is located at a distance of $\dl\sim 8.2$~kpc.  
According to the wide solution, on the other hand, the lens is composed of an object at the 
star/brown-dwarf boundary and an M dwarf with masses $(M_1, M_2)\sim (0.06, 0.33)~M_\odot$ located 
at $\dl\sim 7.7$~kpc.  The binary source is comprised of a primary near the G-dwarf/K-dwarf 
boundary and an early-to-mid M dwarf companion.

\begin{acknowledgements}
Work by C.H. was supported by the grants of National Research Foundation of Korea (2019R1A2C2085965 
and 2020R1A4A2002885).  This research has made use of the KMTNet system operated by the Korea 
Astronomy and Space Science Institute (KASI) and the data were obtained at three host sites of 
CTIO in Chile, SAAO in South Africa, and SSO in
Australia.

\end{acknowledgements}

\end{document}